\documentclass[11pt,a4paper]{article}
\usepackage[top=2.5cm,bottom=2.5cm,left=2.5cm,right=2.5cm]{geometry}
\usepackage[dvipsnames,svgnames,x11names]{xcolor}
\usepackage{amsmath,amssymb,mathtools}
\usepackage{multicol}
\usepackage{titlesec}
\usepackage{fancyhdr}
\usepackage{hyperref}
\usepackage{float}
\usepackage{booktabs}
\usepackage{tabularx}
\usepackage{graphicx}           
\usepackage[section]{placeins}  
\usepackage{caption}            
\usepackage{cuted}

\hypersetup{
    pdfauthor={Tom N. Landefeld, Harald Lesch},
    pdftitle={3D Global Climate Simulations of Synthetic Mars Warming: Assessing the Feasibility Limits of Terraforming},
    hidelinks
}

\titleformat{\section}
  {\normalfont\large\bfseries}{\thesection.}{0.5em}{}
\titlespacing*{\section}{0pt}{1.5ex plus 0.5ex minus 0.2ex}{1ex plus 0.2ex}

\begin{document}

\begin{center}
    \vspace*{2cm}
    {\Large\bfseries 3D Global Climate Simulations of Synthetic Mars Warming: Assessing the Feasibility Limits of Terraforming\par}
    \vspace{2.5em}
    {\large Tom N. Landefeld$^{1}$, Harald Lesch$^{1}$\par}
    \vspace{1.5em}
    {\small $^{1}$University Observatory Munich, Faculty of Physics, Ludwig-Maximilians University of Munich,\par
    Scheinerstraße 1, 81679 Munich, Germany\par}
    \vspace{4.5em}
    {\bfseries Abstract\par}
\end{center}

\vspace{-0.5em}
\noindent
Artificial warming via engineered aerosols presents a novel pathway to bypass volatile scarcity on modern Mars. Utilizing the Generic Planetary Climate Model, we evaluate the physical feasibility of aerosol-induced terraforming over a full Martian year (669 Sols). Unlike previous approaches that added aerosols to background dust or assumed continuous nanoparticle injection, we simulate the complete substitution of natural dust with a single-injection profile of radiatively active aluminium nanoparticles without continuous dumping. This tests whether initial infrared greenhouse forcing can overcome the severe ice-albedo cooling induced by a massive regional water-ice sheet ($3400\text{ kg/m}^{2}$) localized in the Hellas Basin. Comparing this scenario against an unperturbed control run, we analyze surface temperature dynamics, sublimation fluxes, and water vapour concentration. Our 3D global climate simulations quantify the radiative-dynamical feedback thresholds and determine whether engineered aerosol forcing can trigger a self-sustaining water-vapour greenhouse effect or remains fundamentally limited by regional ice-albedo suppression in Hellas.

\vspace{1.5em}
\noindent
\textbf{Keywords:} Mars, climate, Global Climate Model, Engineered aerosols, Nanoparticle radiative forcing, Ice-albedo feedback, Martian water cycle, Hellas Basin, Terraforming

\newpage

\begin{multicols}{2}

\section{Introduction}

Reconstructing the climatic evolution of early Mars remains a major problem in planetary climate science, while extensive geomorphological and mineralogical evidence, such as dendritic valley networks, open-basin lakes, and widespread hydrated silicates, indicates the historical presence of flowing and standing liquid water across ancient Noachian terrain [e.g., Kerber et al., 2015; Wordsworth et al., 2015; Wordsworth, 2016; Haberle et al., 2017; Ojha et al., 2020; Maurice et al., 2026], the consistent reconstruction of geomorphological and mineralogical evidence for persistent or recurrent surface-water activity within the framework of the Faint Young Sun Paradox, when the young Sun emitted approximately 70--75\% of its present luminosity, continues to pose a significant challenge for three-dimensional global circulation models [e.g., Forget et al., 2013; Wordsworth et al., 2013, 2015; Turbet et al., 2019; Ojha et al., 2020; Maurice et al., 2026]. Comprehensive 3D climate simulations using the three-dimensional simulation with the Generic Planetary Climate Model show that even dense $\text{CO}_2$ atmospheres, at surface pressures ranging from 0.1 to 7 bar, do not allow for annual average surface temperatures above the freezing point of water [e.g., Forget et al., 2013; Wordsworth et al., 2013, 2015; Kerber et al., 2015]. From a thermodynamic perspective, this limit is attributable to the adiabatic cooling of rising air masses over elevated terrain, the moderate infrared opacity of carbon dioxide gas, and the increased planetary albedo caused by Rayleigh scattering and backscattering from carbon dioxide ice clouds [e.g., Forget et al., 2013; Wordsworth et al., 2013, 2015; Wordsworth, 2016; Haberle et al., 2017]. Furthermore, under thicker carbon dioxide atmospheres, thermal coupling with the surface intensifies the elevation-dependent cooling, establishing the icy highlands paradigm in which volatile water ice is inexorably advected toward high-altitude cold traps [e.g., Forget et al., 2013; Wordsworth et al., 2013, 2015; Wordsworth, 2016; Ojha et al., 2020]. The volcanic emission of sulphur dioxide, which has long been discussed as an effective warming mechanism, has been shown in 3D GCM simulations to be insufficient to maintain mild climatic conditions. This is due to photolytic reactions that convert sulphur dioxide into optically dense sulphate and sulphur aerosols, whereby the resulting increase in planetary albedo leads to a net cooling of the surface [e.g., Forget et al., 2013; Wordsworth et al., 2013; Kerber et al., 2015; Wordsworth, 2016; Haberle et al., 2017; Turbet et al., 2019]. Similarly, recent calculations using coupled photochemical-climate models have shown that steady-state hydrogen peroxide concentrations are several orders of magnitude too low to have any radiative effect [Maurice et al., 2026]. For water-ice clouds above high-altitude cirrus layers to cause warming, a cloud structure that is physically hardly realistic would be required to prevent net cooling of the Martian surface, high cloud cover of between $75\,\%$ and $100\,\%$, as well as highly specific particle sizes, would be necessary [e.g., Urata and Toon, 2013; Wordsworth et al., 2013; Wordsworth, 2016; Turbet et al., 2020]. Short-term catastrophic events, such as the impact of large meteoroids or massive outflow events, introduce significant quantities of water vapour [e.g., Wordsworth et al., 2013; Haberle et al., 2017; Turbet et al., 2019, 2020]. This condenses into global cloud cover and falls as precipitation, causing the surface to rapidly return to a state of thermal equilibrium below the freezing point [e.g., Wordsworth et al., 2013; Turbet et al., 2019, 2020]. Atmospheres temporarily enriched with reducing gases such as hydrogen and methane can also trigger a temporary warming, however, the resulting rise in temperature accelerates water evaporation, thereby forcing a rapid remigration of the ice into high-altitude reservoirs [e.g., Kite et al., 2017; Wordsworth et al., 2017; Turbet et al., 2019].

\newpage

The current Martian atmosphere, at an atmospheric pressure of 6 mbar, has a low infrared optical depth, meaning that the resulting greenhouse effect of the $\text{CO}_{2}$ atmosphere is limited to around 5 K [e.g., Haberle, 2013; Haberle et al., 2017; Jakosky and Edwards, 2018; Ansari et al., 2024]. Attempts to restore a milder climate by releasing local volatile reservoirs have failed due to significant inventory constraints, with orbital observations confirming that neither the readily accessible $\text{CO}_{2}$ ice nor the deposits adsorbed in the regolith are sufficient to trigger a runaway greenhouse effect [e.g., Haberle et al., 2017; Jakosky and Edwards, 2018; Ansari et al., 2024; Forget et al., 2025]. Consequently, research in the field of planetary engineering has shifted towards altering the radiative balance of planets using artificially produced aerosols, known as nanoparticles [e.g., Ansari et al., 2024; DeBenedictis et al., 2025; Richardson et al., 2025; Kite et al., 2025]. Whilst synthetic super-greenhouse gases such as hydrofluorocarbons require significant quantities of fluorine, an element that is extremely rare in Martian regolith, artificially produced, conductive aerosols made from abundant surface materials exhibit an efficiency per unit mass that is by more than three orders of magnitude higher than that of gaseous alternatives [e.g., Ansari et al., 2024; DeBenedictis et al., 2025; Richardson et al., 2025; Kite et al., 2024]. Conductive aluminium nanorods with an effective length of approximately $9\ \mu\text{m}$ and a 60:1 aspect ratio exploit nanophotonic resonance to strongly interact with upwelling thermal radiation, attempts to close the infrared spectral windows at around $9\ \mu\text{m}$ and $22\ \mu\text{m}$ while forward-scattering solar radiation down to the surface [e.g., Ansari et al., 2024; DeBenedictis et al., 2025; Richardson et al., 2025; Kite et al., 2025]. Three-dimensional transport simulations of aerosol plumes also show that the wind-driven dispersion of these infrared-active particles is directly coupled to their radiative forcing. The simulations further suggest that radiative-dynamic feedbacks promote self-lofting and significantly amplify the meridional Hadley circulation [e.g., Ansari et al., 2024; Richardson et al., 2025]. However, previous 3D GCM studies on warming caused by nanoparticles were based on scenarios that either assumed static, predefined aerosol layers or assumed active, continuous fountaining and injection at rates of 30 to $60~\mathrm{L / s}$ to compensate for losses due to gravitational sedimentation and dry deposition [e.g., Ansari et al., 2024; Richardson et al., 2025]. The subject of this study is a fundamental physical framework, which has not yet been taken into account, for targeted climate-engineering interventions on a planetary scale. Using a Mars adaptation of the three-dimensional Generic Planetary Climate Model, we are investigating a scenario in which the natural background dust is completely replaced by radiatively active aluminium nanoparticles, the key difference being the absence of continuous replenishment or ongoing aerosol injection, to compensate for gravitational settling and dry deposition. To assess the intrinsic thermal stability and feedback limits of this modified atmosphere, we subject the system to a regional thermodynamic test involving the implementation of a massive, locally confined water-ice blanket of $3400\,\mathrm{kg} / \mathrm{m}^2$ within the Hellas Basin, the deepest topographical depression on Mars. Over the course of a fully simulated Martian year (669 sols), we compare this experiment involving the substitution of nanoparticles with an undisturbed control experiment, in which the natural Martian dust was not replaced, under identical boundary conditions for water ice. We quantitatively analyse the evolution of surface temperature, regional sublimation fluxes and the advection of water vapour in the atmosphere, in order to investigate whether the initial infrared greenhouse forcing caused by substituted nanoparticles can overcome the strong regional cooling caused by ice albedo in Hellas Basin and trigger a self-sustaining water vapour greenhouse feedback, or whether the absence of continuous fountaining causes the climate system to succumb to the suppression caused by ice albedo.

\newpage

\end{multicols}
\begin{multicols}{2}

\subsection*{2. Methods}

\subsubsection*{Global Climate Model Architecture and Dynamical Core}

The three-dimensional climate simulations in this study were conducted using the modern Mars adaptation R4130 from March 2026 of the 3D Generic Planetary Climate Model. The model integrates the LMDZ 3D dynamical core, which solves the classical primitive equations of atmospheric meteorology on an Arakawa-C grid using a finite-difference formulation designed to conserve both potential enstrophy and total angular momentum. The global atmospheric domain was discretized onto a horizontal grid resolution of $64 \times 48$ grid points. Vertically, the atmosphere is resolved using 26 hybrid sigma-pressure layers extending from the surface up to the upper atmosphere. Simulations were integrated over a full Martian year of 669 sols with a dynamical timestep calibrated to capture the diurnal cycle and large-scale wave dynamics. The simulations were initalized at northern spring equinox ($L_s = 0^\circ$) using the standard modern Mars cold and dry baseline climate state of the Generic PCM. The initial atmospheric composition was set to a $CO_2$-dominated atmosphere

($95.32\,\%\text{ CO}_2$, $2.7\,\%\text{ N}_2$ and $1.6\,\%\text{ Ar}$) with a global mean surface pressure (6.1 mbar). Initial atmospheric temperature profiles, surface temperature, and subsurface soil thermal layers were extracted from a converged Mars spin-up state.

\subsubsection*{Physical Parameterizations and Surface Boundary Conditions}

Subgrid-scale dynamical processes including boundary layer turbulence and convective mixing were parameterized using the time-dependent Mellor and Yamada 2.5-level closure scheme, complemented by a moist convective adjustment scheme (calladj=.true., moistadjustment=.true.) that rapidly relaxes unstable vertical temperature profiles to a stable lapse rate.

\vspace{2em}

Sensible heat exchanges between the surface and the first atmospheric layer ($\sim 18\text{ m}$) were calculated using bulk aerodynamic formulations incorporating von K\'arm\'an surface roughness parameters ($z_0 = 0.01\text{ m}$). Subsurface thermal conduction was evaluated across an 18-layer regolith model extending to a depth of several meters, assuming a baseline soil thermal inertia of $250\text{ J m}^{-2}\text{ s}^{-1/2}\text{ K}^{-1}$ and a bare ground albedo of $A_{\text{rock}} = 0.20$. The global water cycle was fully activated (water=.true.), incorporating dynamic water-ice cloud condensation (watercond=.true.) radiatively active water-ice clouds (aeroh2o=.true.) with a fixed effective ice particle radius of $r_{\text{h2o\_ice}} = 35\,\mu\text{m}$, cloud fraction variability (CLFvarying=.true.), precipitation processes (waterrain=.true., precip\_scheme=1), surface hydrology (hydrology=.true.), and evolving surface water sources (sourceevol=.true.). To impose a severe regional thermodynamic test, a localized, massive water-ice sheet of $3400\text{ kg/m}^2$ with a surface albedo of $A_{\text{ice}} = 0.55$ was placed in the Hellas Basin $40^\circ\text{S} - 60^\circ\text{S}$.

\subsubsection*{Radiative Transfer and Nanoparticle Substitution Setup}

Radiative transfer was computed using the correlated-k distribution method, utilizing 40 spectral bands in the thermal infrared and 38 bands in the visible spectrum. Molecular gas opacities for $\text{CO}_2$ and $\text{H}_2\text{O}$ were incorporated alongside collision-induced absorption continua. In the experimental run, natural Martian background dust was completely replaced by a single-injection with radiatively active aluminium nanoparticles. Unlike continuous fountaining setups, no active aerosol injection or ongoing replenishment was applied during the simulation (timedepdust = .true.). The optical properties of the aluminium particles (material density $\rho_{\text{part}} = 2700\text{ kg/m}^3$) including mass extinction cross-sections, single-scattering albedos, and asymmetry parameters were tabulated across all model spectral bands to account for resonant infrared extinction and forward-scattering in the visible spectrum.
\end{multicols}

\newpage
\subsection*{3. Results}

\begin{multicols}{2}

The numerical simulation over a period of 669 sol (1 Martian year)
using the 3D global circulation model Generic PCM reveals marked
differences between the undisturbed Martian background climate and
the scenario in which the background dust is completely replaced by
aluminium nanoparticles. The physical-diagnostic analysis of the
state variables reveals pronounced deviations in the thermodynamic,
radiative and hydrological regimes of the climate system.

\vspace{1em}

\noindent
\includegraphics[width=\linewidth]{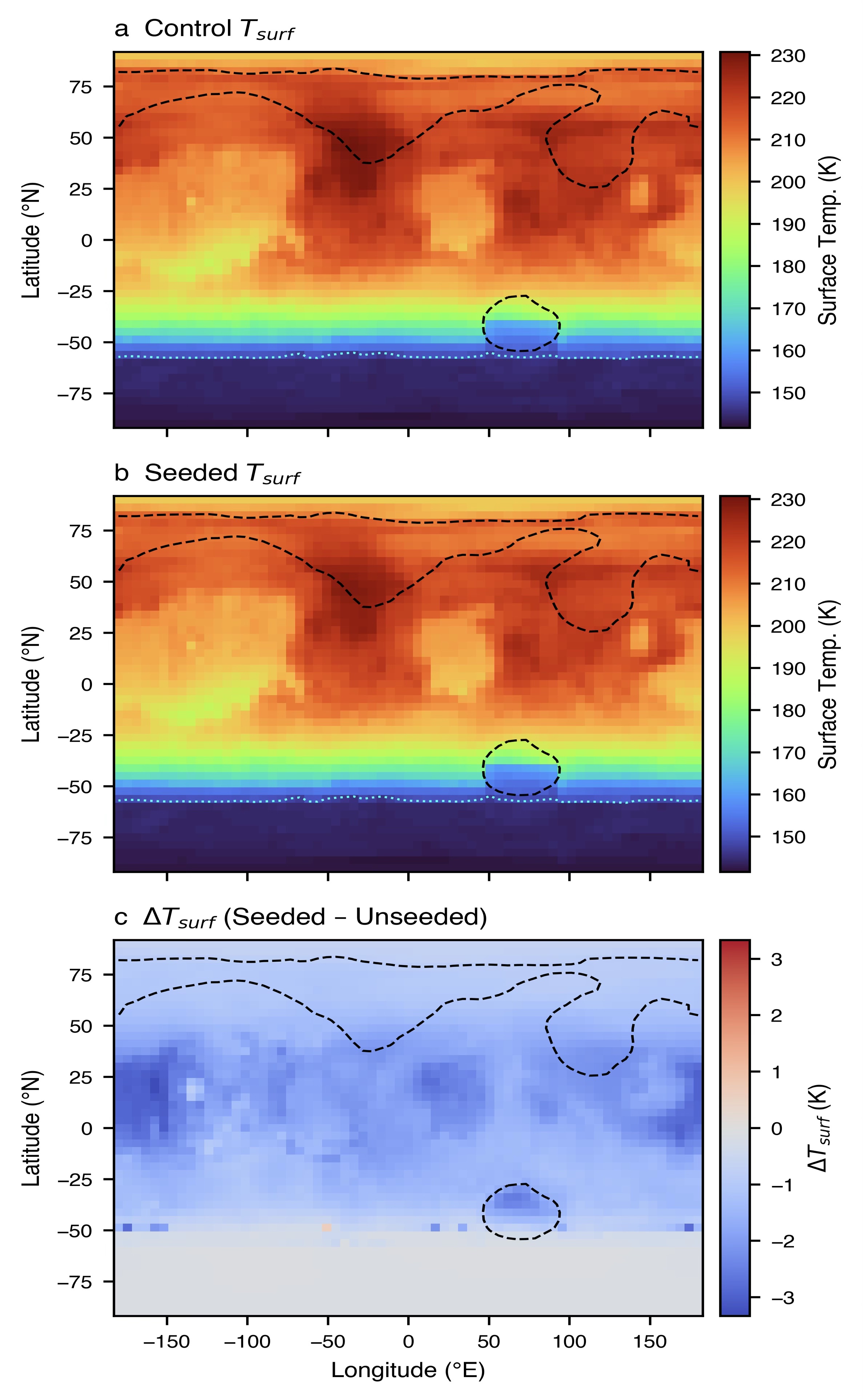}

\vspace{2mm}

{\footnotesize
\setlength{\parfillskip}{0pt plus 1fil}
\textbf{Figure 1.} Global mean surface temperature $T_{\text{surf}}$. 
(a) Annual mean surface temperature for the control simulation. 
(b) Annual mean surface temperature under single-injection nanoparticle substitution. 
(c) Surface temperature anomaly $\Delta T_{\text{surf}}$, illustrating suppressed surface heating directly over the high-albedo Hellas Basin ice sheet.\par}

\columnbreak

{\large\textbf{Thermodynamic state and surface albedo}}

\vspace{0.8em}

The nanoparticle simulation shows a systematic decrease in the
global mean surface temperature $T_{\text{surf}}$ (Figure 1b) to
$191.80\,\text{K}$, compared with $194.89\,\text{K}$ in the
control run (Figure 1a) under conditions with the dust
(Extended Data Tables 1 and 2). This corresponds to an average
global cooling of the surface by
$\Delta T_{\text{surf}} = -3.09\,\text{K}$ (Figure 1c).
The seasonal and diurnal temperature maximum decreases from
$313.22\,\text{K}$ in the control run to $304.07\,\text{K}$
in the nanoparticle scenario, representing an attenuation of
the thermal peaks by $9.15\,\text{K}$.

The spatial distribution of surface albedo ($ALB$) (Figure 2)
shows only minor differences (Figure 2c), on a global average,
between the seeded run (0.1348) and the control run (0.1375).
Regional albedo consistency is largely attributable to the
optical properties of the uncovered basalt ground, whereas
local asymmetries vary in correlation with the dynamics of
the seasonal polar ice caps.

\end{multicols}
 
 \newpage

\begin{multicols}{2}

\begin{center}
  \includegraphics[width=\linewidth]{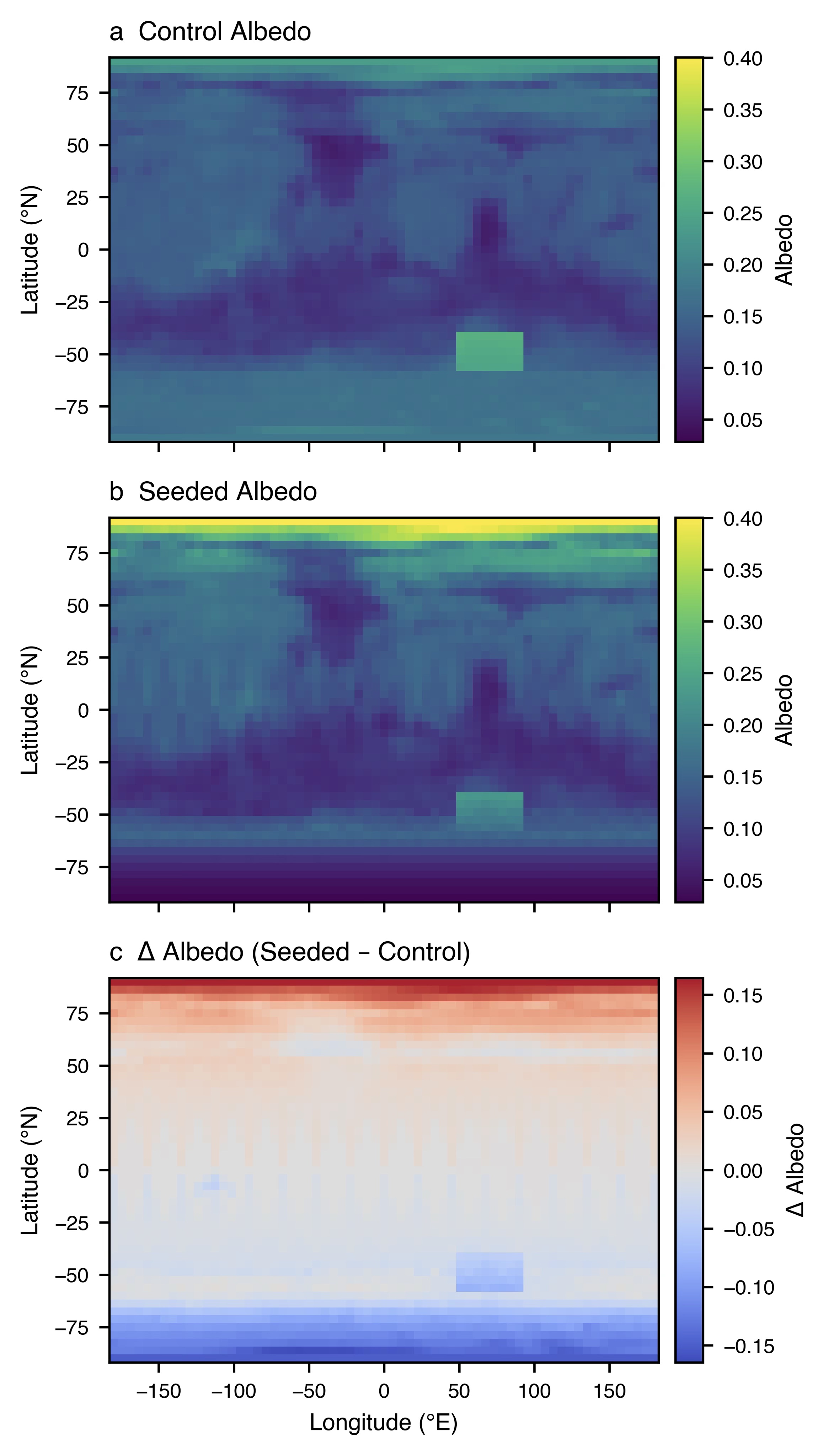}
\end{center}

\vspace{-2mm}
{\small\textbf{Figure 2.} Mean Surface albedo distribution\par
\textbf{(a)} Global surface $Albedo$ for the unperturbed control run \textbf{(b)} Global surface $Albedo$ for the nanoparticle substitution simulation \textbf{(c)} Albedo anomaly $\Delta Albedo$, highlighting the persistent high-albedo ($A = 0.55$) of the Hellas Basin ice sheet ($40^\circ\text{S} - 60^\circ\text{S}$, $50^\circ\text{E} - 100^\circ\text{E}$) alongside localized marginal shifts due to reduced seasonal frost accumulation.\par}

\subsubsection*{Vertical thermal structure of the atmosphere}

Atmospheric temperatures $T_{\text{atmos}}$ deoicted in Figure 3 correspond to the near-surface air temperature, providing a direct metric for surface-atmosphere sensible heat exchange. The decrease in the near-surface atmospheric temperature $T_{\text{atmos}}$ is more pronounced than at the surface. The mean atmospheric temperature falls from $176.16\text{ K}$ in the control run to $169.15\text{ K}$ in the nanoparticle scenario (seeded). The diagnosed atmospheric temperature maximum decreases from $273.29\text{ K}$ in the control run to $250.39\text{ K}$ in the seeded run, indicating substantially colder atmospheric conditions throughout the simulated year.

\begin{center}
  \includegraphics[width=\linewidth]{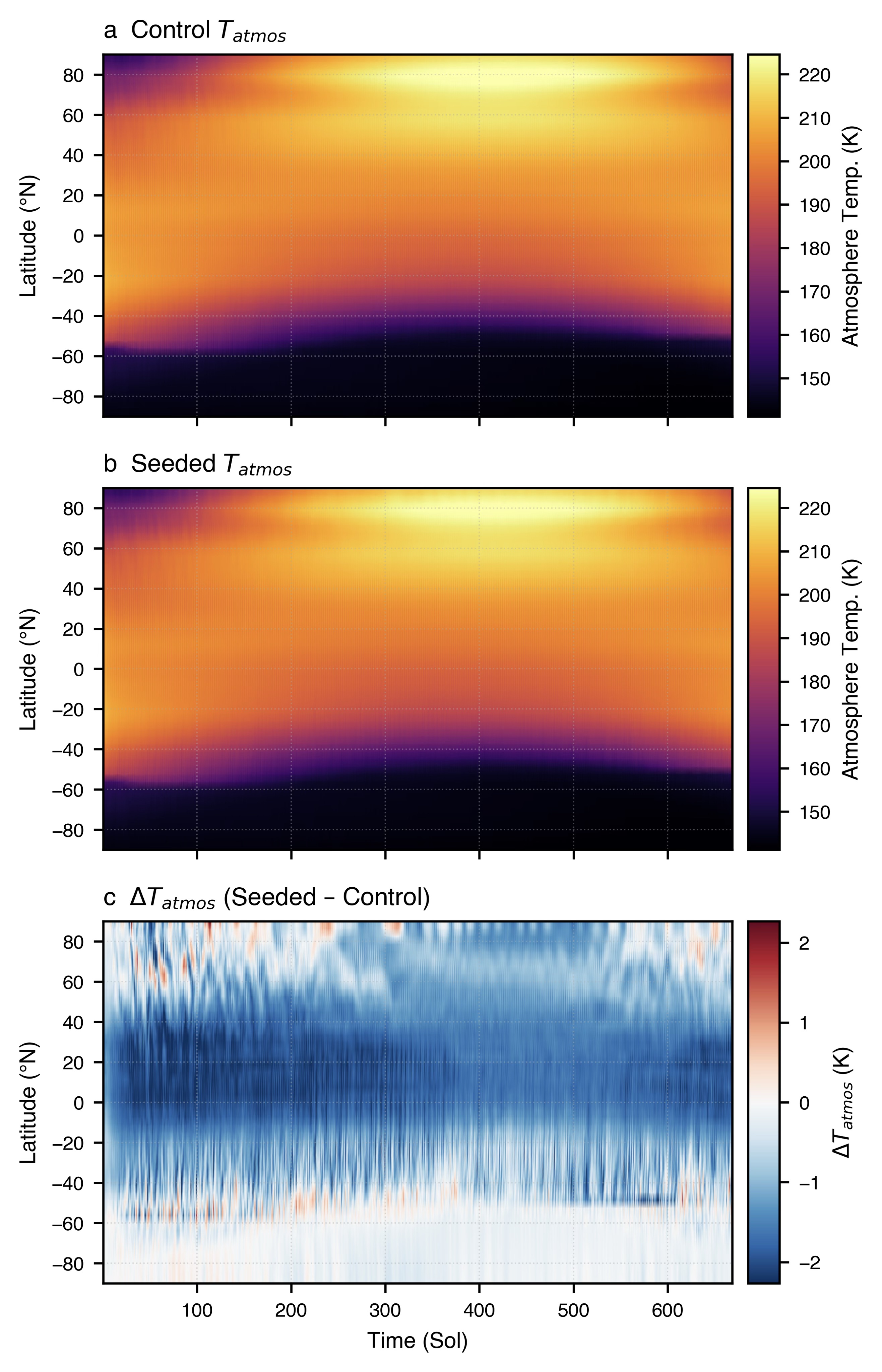}
\end{center}

\vspace{-2mm}
{\small\textbf{Figure 3.} Seasonal evolution of near surface atmospheric temperature $T_{\text{atmos}}$ over (669 Sol)\par
\textbf{(a)} Atmospheric temperature structure for the control simulation \textbf{(b)} Atmospheric temperature evolution under single-injection aluminium nanoparticle substitution \textbf{(c)} Atmospheric temperature anomaly $\Delta T_{\text{atmos}}$, demonstrating the transient nature of the radiative forcing and the decay of atmospheric warming over time due to aerosol sedimentation in the absence of continuous replenishment.\par}

\subsubsection*{Atmospheric mass, pressure and $\text{CO}_2$-phase equilibrium}

The thermal depression triggers a large-scale redistribution of the main volatile atmospheric component. As a result of the reduced atmospheric temperatures, the phase equilibrium of $\text{CO}_2$ shifts in favour of the solid phase. The mean global surface pressure $Pa_{\text{surf}}$ falls from $515.78\text{ Pa}$ to $452.38\text{ Pa}$. This corresponds to a decrease in the total atmospheric mass of $12.29\,\%$ due to condensation.

\end{multicols}

\begin{center}
 
\end{center}

\begin{multicols}{2}

\begin{figure}[H]
\centering
\includegraphics[width=\linewidth]{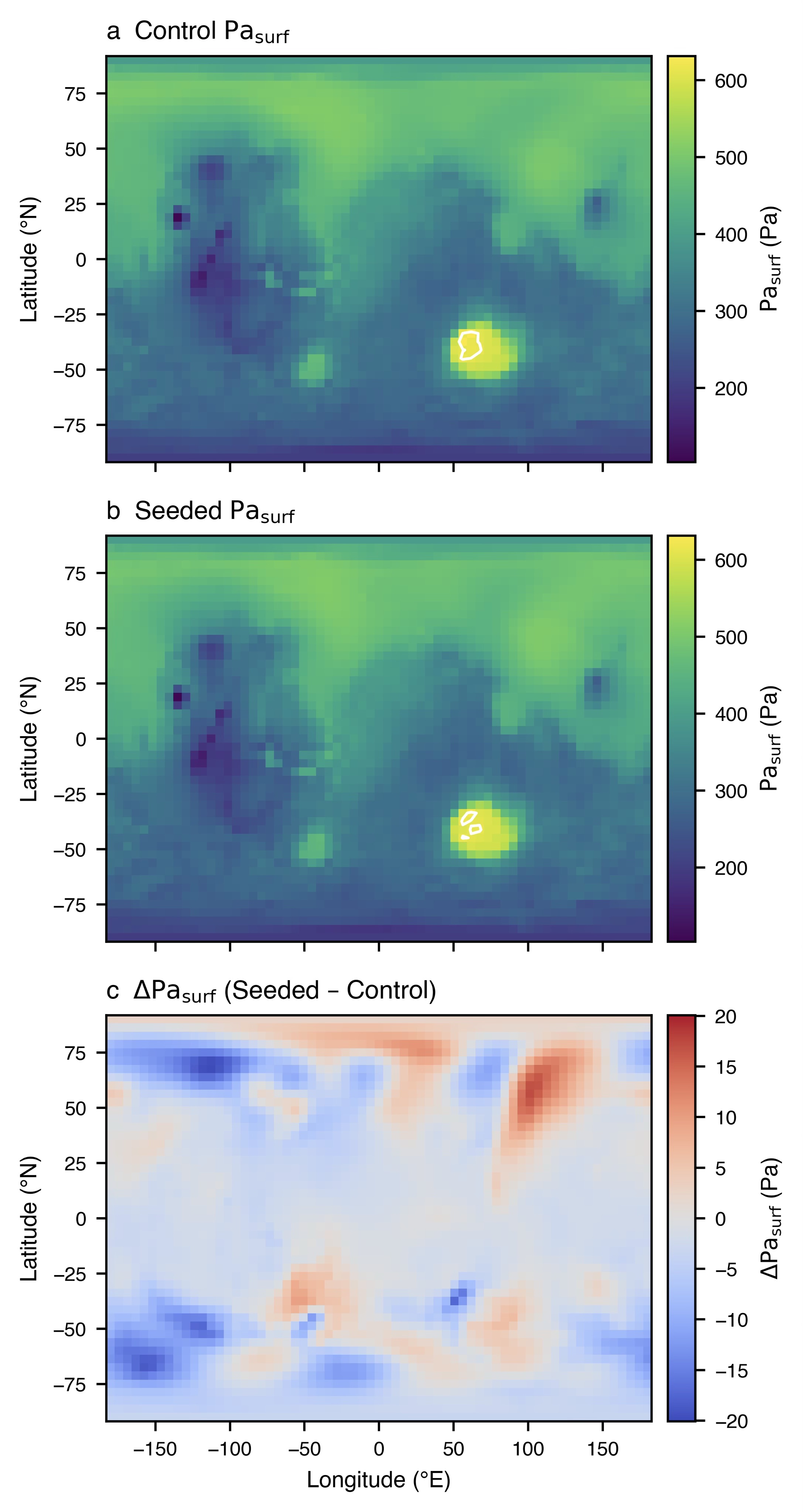}
\end{figure}

\vspace{-1em}
\textbf{Figure 4.} Surface pressure distribution and topographic control\\
\textbf{(a)} Mean surface pressure ($Pa_{surf}$) for the control simulation, illustrating the hypsometric pressure maximum within the Hellas Basin \textbf{(b)} Surface pressure field under single-injection nanoparticle substitution \textbf{(c)} Net pressure anomaly $\Delta Pa_{surf}$ confirming the absence of a global pressure expansion

\vspace{0.5em}
In contrast to this reduction, the mass per unit area of the deposited dry ice on the surface increases, on a global annual average, from $180.38\text{ kg/m}^2$ in the control run to $223.30\text{ kg/m}^2$ in the seeded run, which corresponds to an increase of $+23.79\,\%$ (Extended Data Figure 1).

\subsection*{Sublimation dynamics and ice inventories}
Despite the presence of massive local water-ice reservoirs of up to $3400\text{ kg/m}^2$ (Extended Data Figure 4) in the deep Hellas Basin, the hydrological cycle in the seeded scenario shows an almost complete collapse. The global mean water vapour mixing ratio $q_{\text{vap}}$ falls by more than two orders of magnitude from to $3.08 \times 10^{-7}\text{ kg/kg}$ (a decrease of $99.65\,\%$). The maximum water mixing ratio plummets from $0.207\text{ kg/kg}$ to $1.33 \times 10^{-3}\text{ kg/kg}$.

\begin{figure}[H]
\centering
\includegraphics[width=\linewidth]{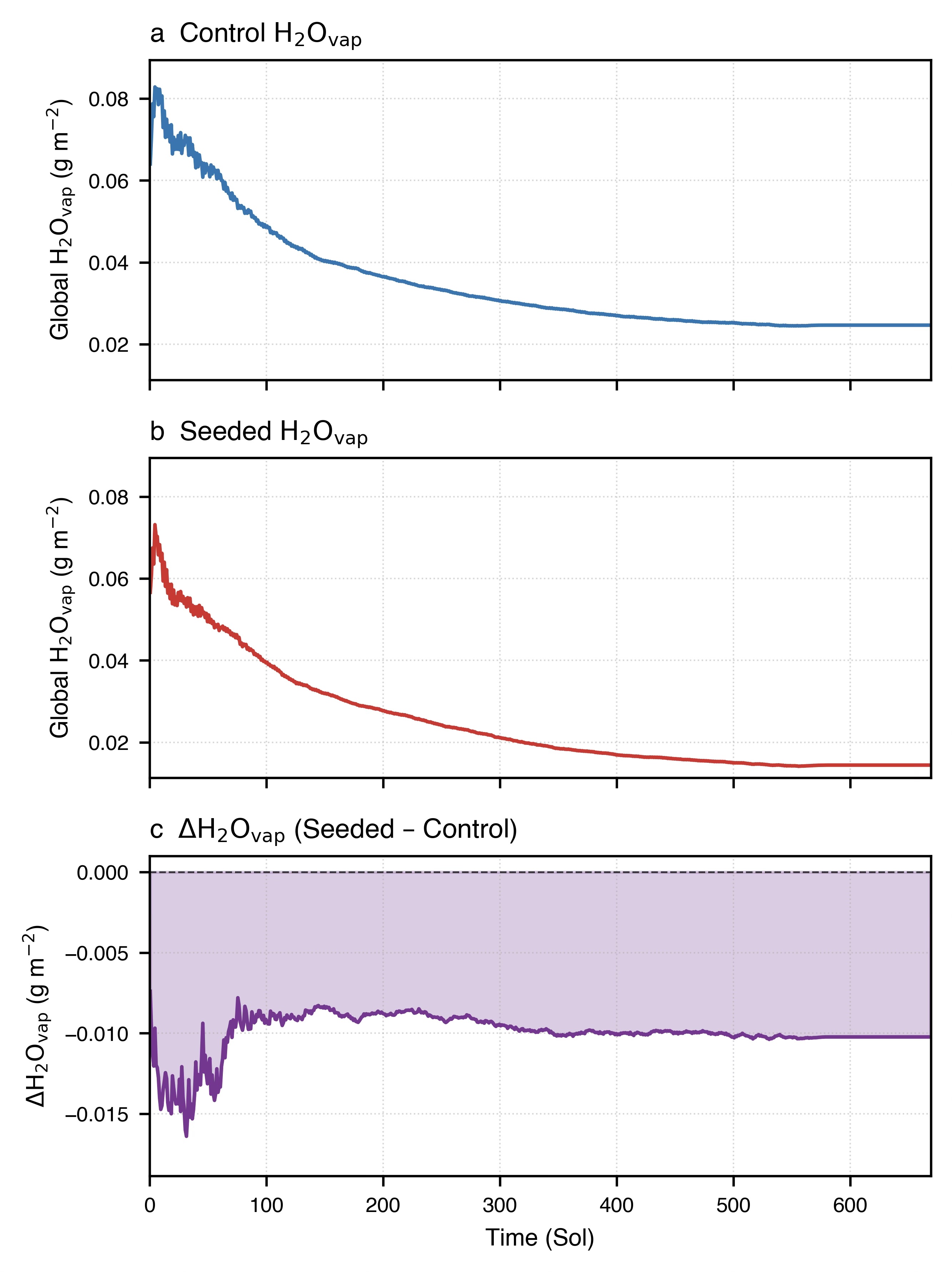}
\end{figure}

\vspace{-1em}
\textbf{Figure 5.} Atmospheric water vapour evolution across 669 Sol\\
\textbf{(a)} Column-integrated water vapour abundance $\mathrm{H_2O_{vap}}(\mathrm{g/m}^2)$ for the control run\\
\textbf{(b)} Column-integrated water vapour abundance under single-injection nanoparticle substitution \textbf{(c)} Water vapour anomaly $\Delta\mathrm{H_2O_{vap}}$, documenting a persistent global reduction in atmospheric water vapour

\newpage

\vspace{1em}
This dehydration of the atmosphere correlates directly with the suppression of the latent heat flux and the sublimation rates (Extended Data Figure 2). Whilst maximum sublimation rates of $169.12\text{ W/m}^2$ are reached in the control run, the peak values in the seeded run are reduced to just $0.374\text{ W/m}^2$ (a drop of $99.78\,\%$). It is important to emphasize that this trajectory does not represent a stationary, recurring seasonal equilibrium cycle, but rather a transient relaxation path towards a dehydrated steady state. At initialization ($\text{Sol} = 0$), the prescribed initial conditions and the localized Hellas Basin water-ice reservoir support elevated initial atmospheric moisture levels in both setups. However, while the cold background climate in the control run rapidly suppresses evaporation and causes vapour freeze-out, the continuous gravitational sedimentation of nanoparticles in the perturbed run additionally dismantles radiative warming over the course of the Martian year. As radiative warming collapses, atmospheric temperatures drop below the condensation threshold, forcing rapid freeze-out of atmospheric moisture. By the end of Year 1 ($L_{\mathrm{s}} \approx 360^{\circ} / \text{Sol} = 669$), atmospheric water vapour in both runs drops to a near zero floor. Multi-year extensions of this setup confirm that the system remains locked at this asymptotic minimum throughout subsequent years without seasonal recovery, demonstrating that full thermodynamic and hydrological equilibrium is achieved at the end of Year 1. Consequently, depicting additional simulation years is redundant, as it yields a flat zero-line with no further physical dynamics. As a result, atmospheric water-ice condensation is also reduced. The vertically integrated water ice column density (Extended Data Figure 3), which describes the atmospheric ice mass and must be strictly distinguished from optical depths, decreases on an annual average from $1.89 \times 10^{-6}\text{ kg/m}^2$ (maximum $4.12 \times 10^{-3}\text{ kg/m}^2$) to $6.15 \times 10^{-9}\text{ kg/m}^2$ (maximum $8.97 \times 10^{-8}\text{ kg/m}^2$). The simulation results thus demonstrate a pronounced depression accompanied by a simultaneous decrease in aerosol optics and atmospheric water vapour content, whilst the $\text{CO}_2$ ice inventory increases. 

\vspace{3em}

The physical dynamics can be understood as a coupled chain of effects comprising aerosol loss, radiative cooling and suppression of the hydrological cycle, although the individual quantitative contributions cannot be completely disentangled on the basis of the available data set.

\subsection*{Aerosol optics, column density and sedimentation kinematics}

The optical diagnostics show reduced atmospheric opacity in the seeded run. The mean total opacity $\tau_{\text{col}}$ (Extended Data Figure 5) falls from $0.197$ to $0.141$, whilst the specific aerosol opacity $\tau_{\text{dust}}$ decreases on average from $0.164$ (maximum $0.773$) to $0.112$ (maximum $0.322$) (Extended Data Figure 6). It should be emphasised that the optical depth $\tau_{\text{dust}}$ is an integrated radiative diagnostic quantity and that the underlying sedimentation kinematics must not be derived solely from the change in opacity $\Delta\tau$. Physically, the behaviour of the nanoparticles at atmospheric pressures of $ps_{\text{atmos}} \approx 450\text{--}515\text{ Pa}$ requires a kinetic classification of the flow regime. With a mean free path length of the $\text{CO}_2$ molecules of $\lambda_{\text{fp}} \approx 10\,\mu\text{m}$, the Knudsen number $Kn = \frac{\lambda_{\text{fp}}}{r_{\text{part}}}$ lies in the range $Kn \approx 10^3 \gg 1$.

The nanoparticle aerosol is therefore unambiguously in the free-molecule regime. In this kinetic regime, momentum transfer is dominated by discrete molecular collisions, which results in a significant reduction in frictional resistance compared with the continuum regime. The Cunningham-corrected Stokes' approach for the sedimentation velocity
\begin{equation}
v_{\text{settling}} = \frac{2 \cdot \rho_{\text{part}} \cdot r_{\text{part}}^2 \cdot g_{\text{mars}} \cdot C_{\text{c}}}{9 \cdot \mu_{\text{air}}}
\end{equation}
formally transitions, for $Kn \gg 1$ (where $C_{\text{c}} \approx A \cdot Kn$), to the Epstein settling velocity, which depends linearly on the particle radius
\begin{equation}
v_{\text{settling,Epstein}} \propto \frac{\rho_{\text{part}} \cdot r_{\text{part}}}{\rho_{\text{gas}} \cdot v_{\text{th}}}.
\end{equation}

The lower opacity diagnosed in the model therefore reflects the interplay between kinetic deposition in the free-molecule regime, atmospheric transport processes and the absence of a continuous fountaining source.

\end{multicols}

\begin{multicols*}{2}[\subsection*{4. Discussion}]
\noindent
The three-dimensional global circulation simulations using the Generic PCM model demonstrate that the complete substitution of natural background dust with aluminium nanoparticles without fountaining is insufficient to initiate self-sustaining greenhouse warming, provided that the climate system is constrained by a regional water-ice reservoir ($3400\text{ kg/m}^2$, ($A = 0.55$)) in the Hellas Basin. Although aluminium nanorods exhibit high extinction in the thermal infrared in addition to plasmonic resonances, a significant proportion of the incident short-wave solar radiation is reflected due to the high surface albedo of the ice sheet. In the single-injection substitution simulation without continuous injection, the global mean surface temperature ($T_{\mathrm{surf}}$) decreases from $194.89\text{ K}$ in the control run to $191.80\text{ K}$ in the nanoparticle scenario ($\Delta T_{\mathrm{surf}} = -3.09\text{ K}$). Correspondingly, the seasonal and diurnal temperature maxima decrease from $313.22\text{ K}$ to $304.07\text{ K}$ ($\Delta T_{\mathrm{surf}} = 9.15\text{ K}$). This local thermal depression over the Hellas Basin significantly suppresses both the latent heat fluxes and the sublimation rates. The result is global dehydration of the atmosphere, with the column mixing ratio of water vapour falling from $8.82 \times 10^{-5}\text{ kg/kg}$ to $3.08 \times 10^{-7}\text{ kg/kg}$, whilst the accumulation of solid $\mathrm{CO}_2$ ice on the surface rises from a global average of $180.38\text{ kg/m}^2$ to $223.30\text{ kg/m}^2$. As the low density of the atmosphere causes the volatile water to condense in cold traps, the positive water vapour greenhouse feedback does not occur. In summary, the three-dimensional climate simulation demonstrates neither the unrestricted functionality nor the outright failure of planetary geoengineering concepts. The results show that the artificial warming of Mars using nanoparticles constitutes a transient radiative pulse, the climatic efficacy of which is strictly limited by interactions with regional ice-albedo feedbacks, the suppression of the hydrological cycle, and dynamically determined sedimentation timescales. A sustainable modification of the climate system therefore requires far more than an initial dust substitution. 

\vspace{3em}

Its effectiveness depends crucially on the maintenance of sufficient atmospheric aerosol optics, as well as on the interaction between aerosol residence time, surface ice and the atmospheric water cycle.

\end{multicols*}

\begin{minipage}[t]{0.47\textwidth}
\subsection*{\large\textbf{5. Extended Data}}
\vspace{0.8em}
{\large\textbf{Extended Data Figure 1}}
\vspace{0.6em}

\includegraphics[width=\linewidth]{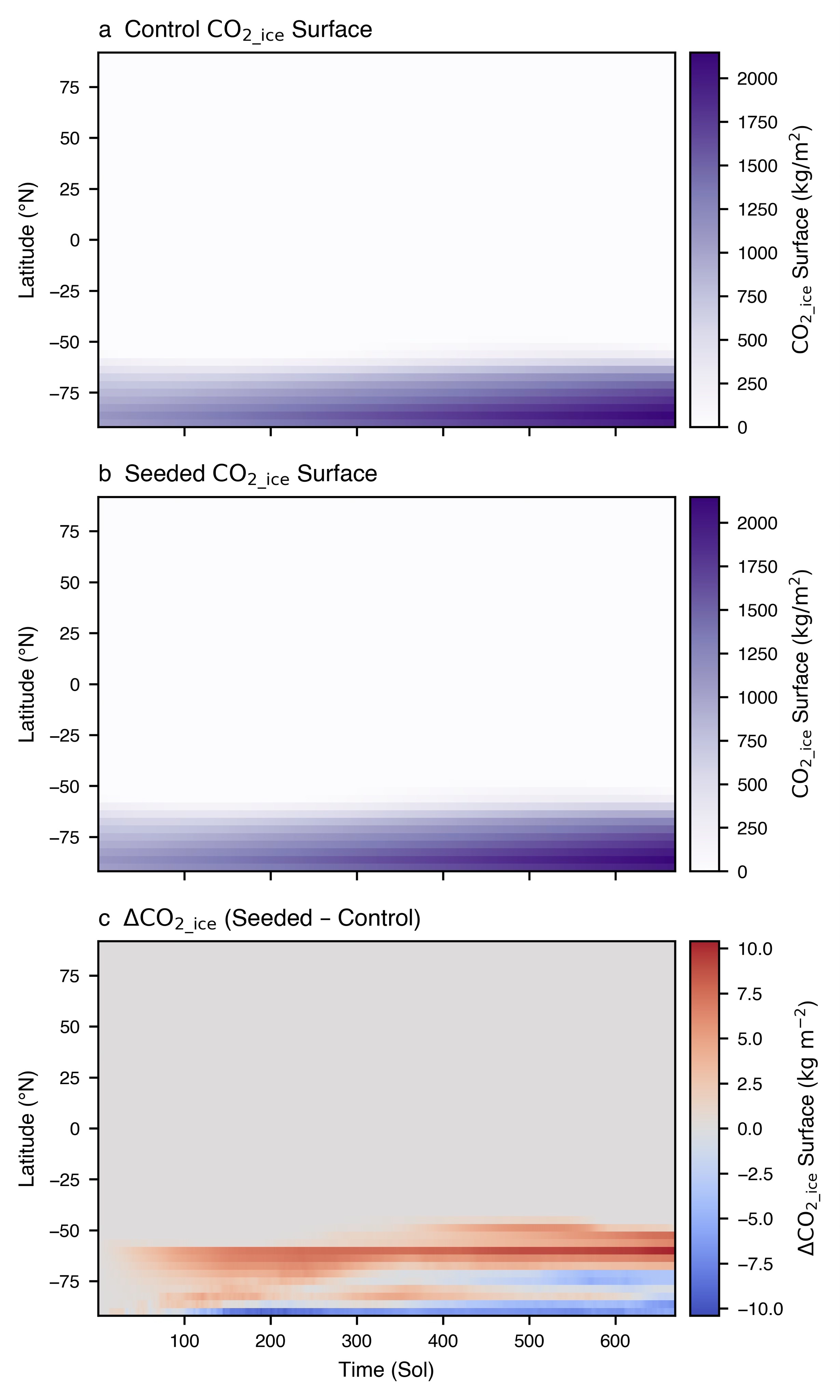}

\vspace{0.8em}
\textbf{Extended Data Figure 1.} Seasonal evolution of surface $\text{CO}_{2}$ ice across 669 Sol

\textbf{(a)} Surface $\text{CO}_2$ ice mass distribution for the control simulation, showing the seasonal accumulation of $\text{CO}_2$ ice at high southern latitudes \textbf{(b)} Surface $\text{CO}_2$ ice distribution under single-injection aluminium nanoparticle substitution \textbf{(c)} Difference in surface $\text{CO}_2$ ice mass between the seeded and control simulations, showing increased $\text{CO}_2$ ice accumulation in the seeded scenario
\end{minipage}%
\hfill
\begin{minipage}[t]{0.47\textwidth}
\vspace*{2.35em}
{\large\textbf{Extended Data Figure 2}}
\vspace{0.6em}

\includegraphics[width=\linewidth]{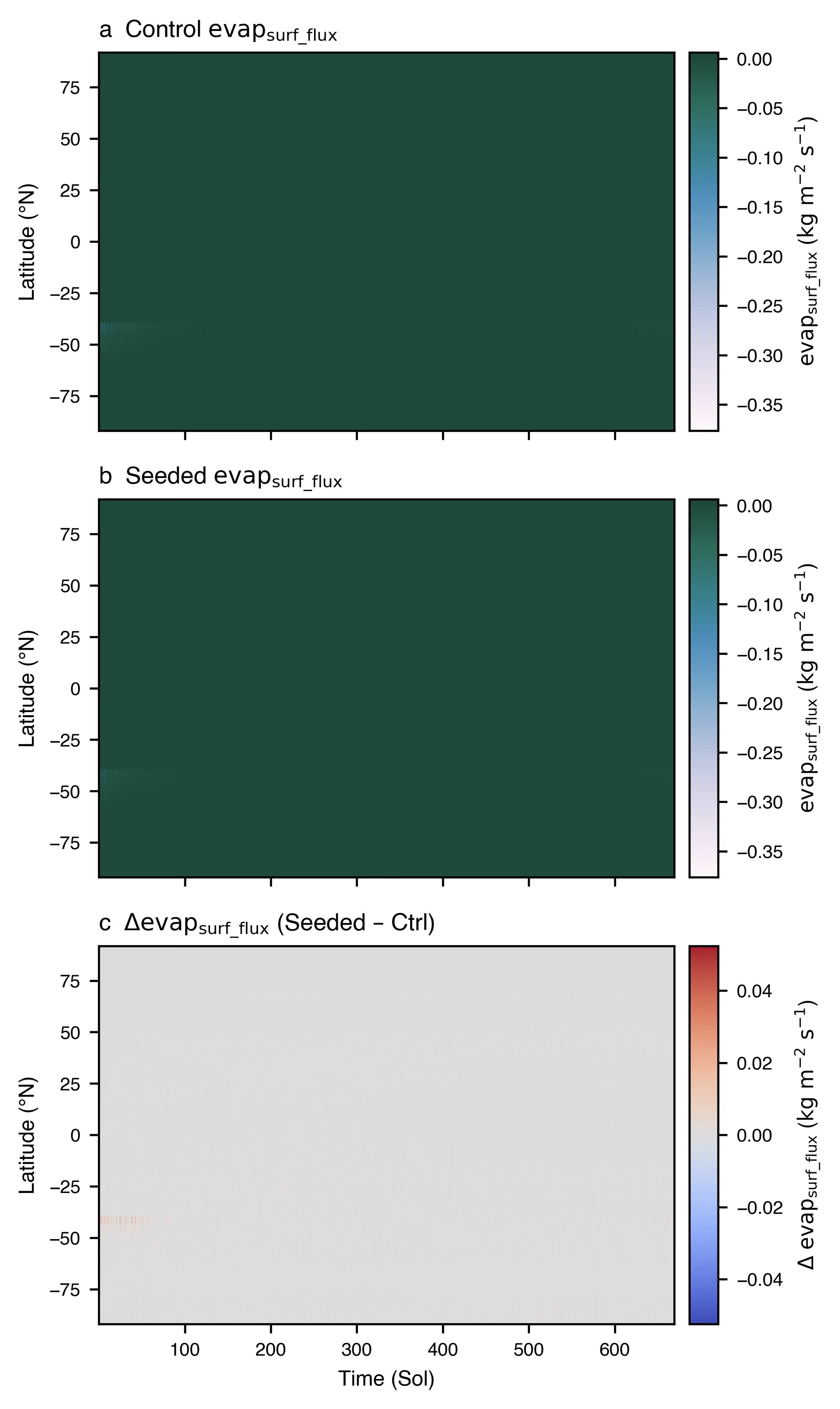}

\vspace{0.8em}
\textbf{Extended data Figure 2.} Seasonal evolution of surface water flux across 669 Sol

\textbf{(a)} Surface water flux in the control simulation \textbf{(b)} Surface water flux under single-injection aluminium nanoparticle substitution, showing strongly reduced fluxes \textbf{(c)} Surface water flux anomaly between the seeded and control simulations, documenting the suppression of surface water exchange under nanoparticle forcing
\end{minipage}

\vfill
\begin{center}

\end{center}

\newpage

\begin{minipage}[t]{0.47\textwidth}
{\large\textbf{Extended Data Figure 3}}
\vspace{0.8em}

\includegraphics[width=\linewidth]{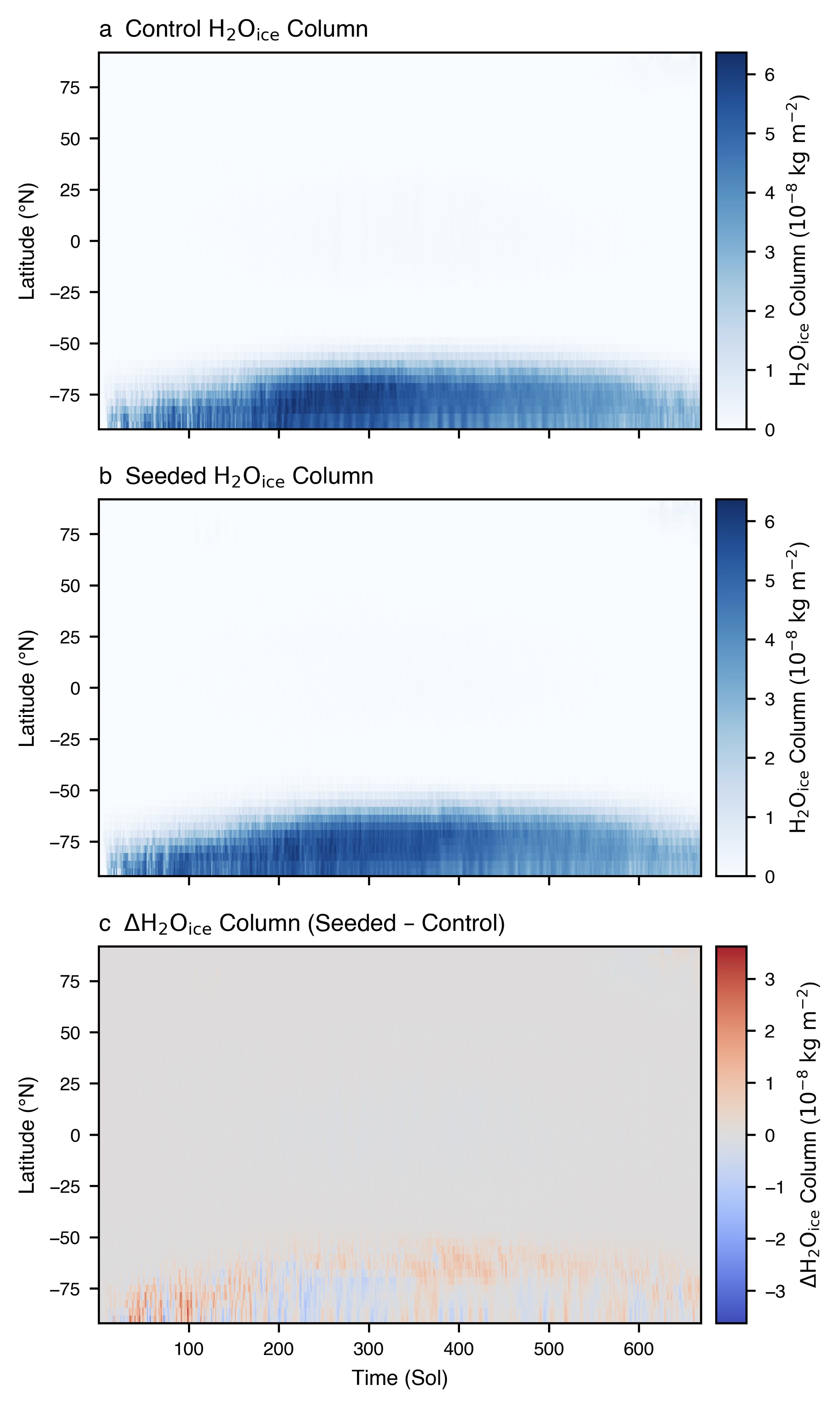}

\vspace{0.8em}
\textbf{Extended Data Figure 3.} Atmospheric water-ice column density across 669 Sol

\textbf{(a)} Vertically integrated atmospheric water-ice column density for the control simulation \textbf{(b)} Atmospheric water-ice column density under single-injection aluminium nanoparticle substitution \textbf{(c)} Difference in atmospheric water-ice column density between the seeded and control simulations, showing the reduced atmospheric water-ice inventory in the seeded scenario
\end{minipage}%
\hfill
\begin{minipage}[t]{0.47\textwidth}
{\large\textbf{Extended Data Figure 4}}
\vspace{0.8em}

\includegraphics[width=\linewidth]{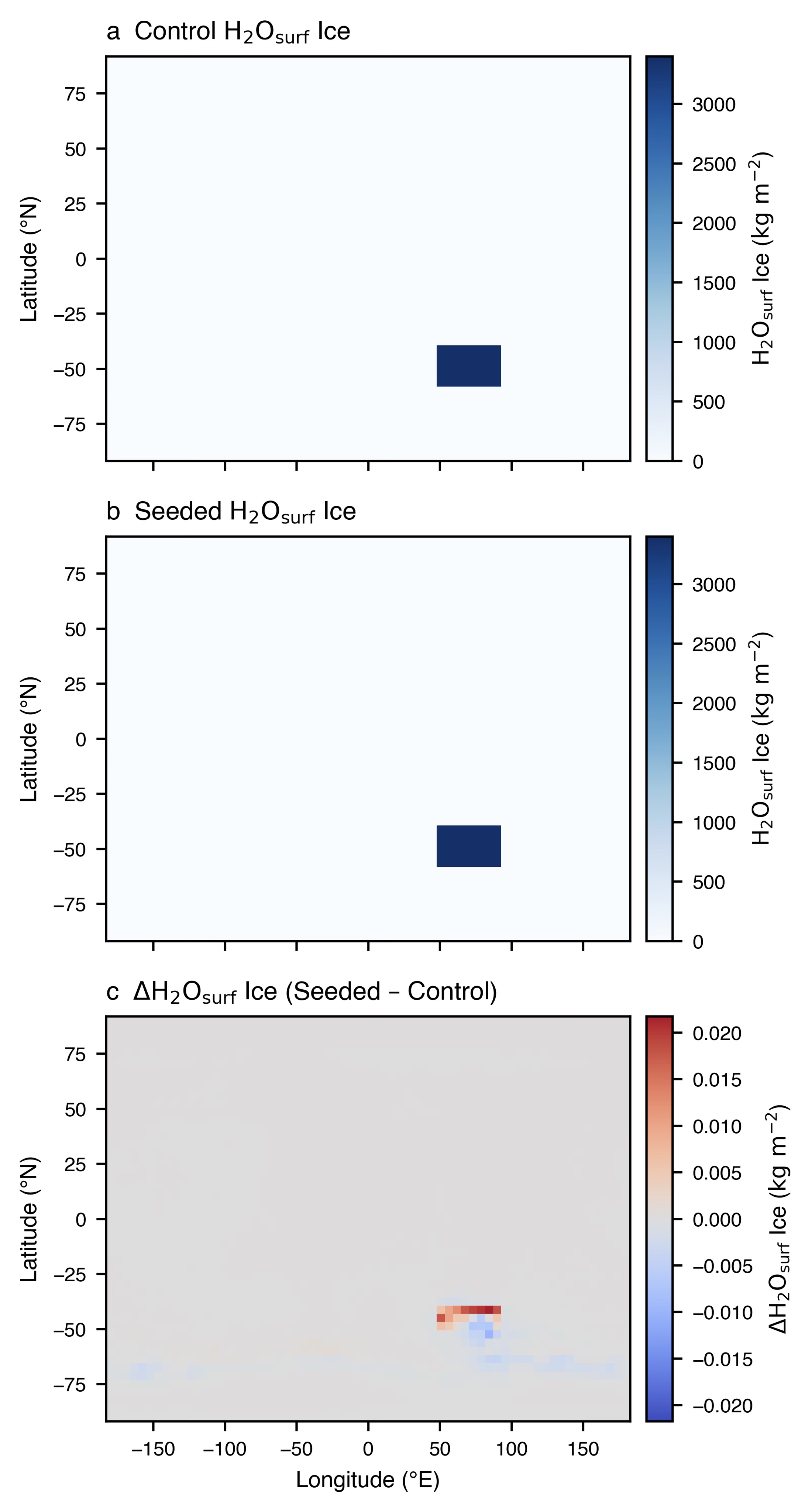}

\vspace{0.8em}
\textbf{Extended Data Figure 4.} Surface water-ice source distribution

\textbf{(a)} Surface water-ice source distribution for the control simulation, showing the imposed water-ice reservoir in the Hellas Basin \textbf{(b)} Surface water-ice source distribution under single-injection aluminium nanoparticle substitution \textbf{(c)} Anomaly in surface water-ice source between the seeded and control simulations, showing the localized change associated with the imposed ice reservoir
\end{minipage}

\vfill
\begin{center}

\end{center}

\newpage

\begin{minipage}[t]{0.47\textwidth}
{\large\textbf{Extended Data Figure 5}}
\vspace{0.8em}

\includegraphics[width=\linewidth]{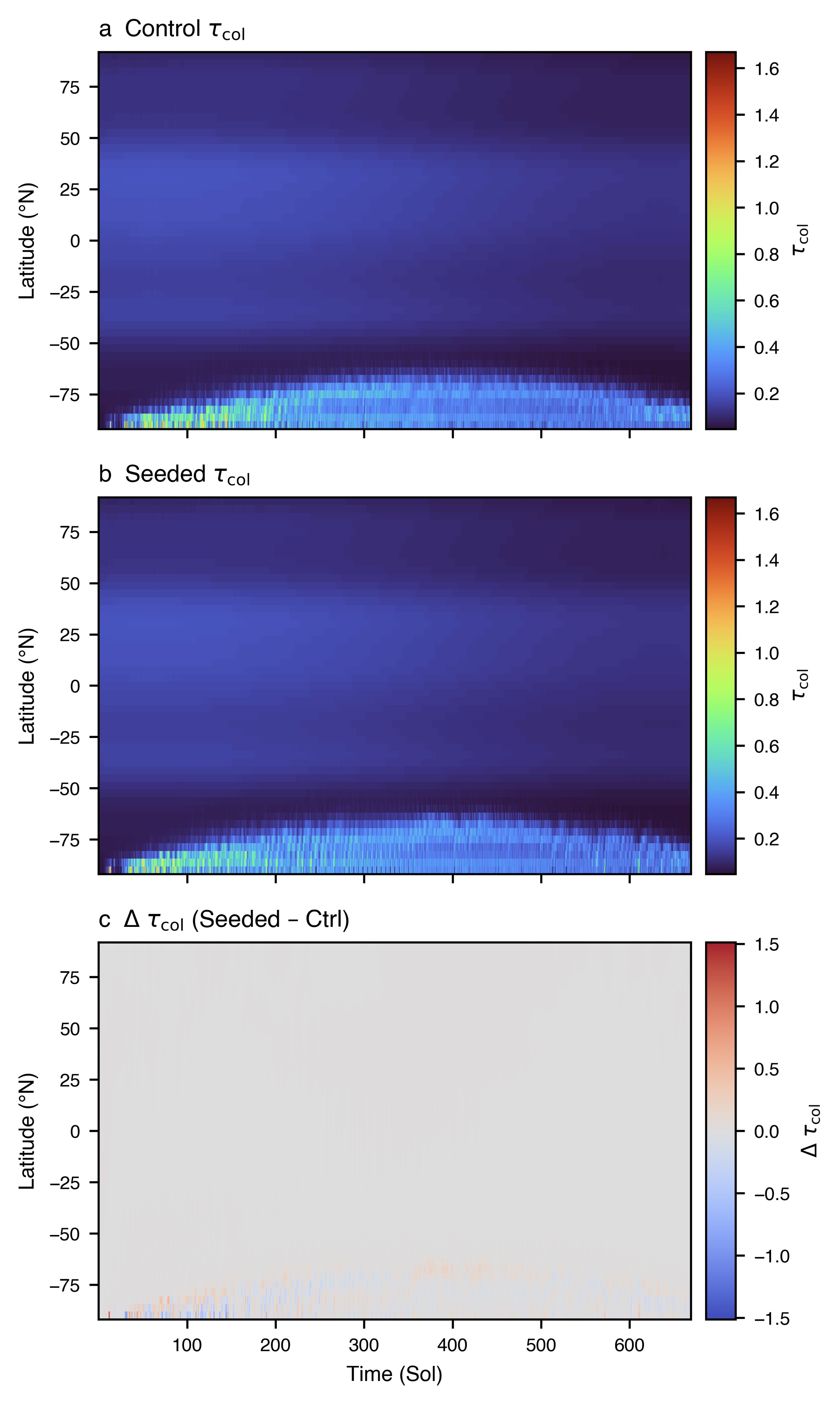}

\vspace{0.8em}
\textbf{Extended Data Figure 5.} Seasonal evolution of total atmospheric optical depth across 669 Sol

\textbf{(a)} Total atmospheric optical depth for the control simulation \textbf{(b)} Total atmospheric optical depth under single-injection aluminium nanoparticle substitution \textbf{(c)} Total optical-depth anomaly between the seeded and control simulations, showing the temporal evolution of the radiative effect associated with the aerosol substitution
\end{minipage}
\hfill
\begin{minipage}[t]{0.47\textwidth}
{\large\textbf{Extended data Figure 6}}
\vspace{0.8em}

\includegraphics[width=\linewidth]{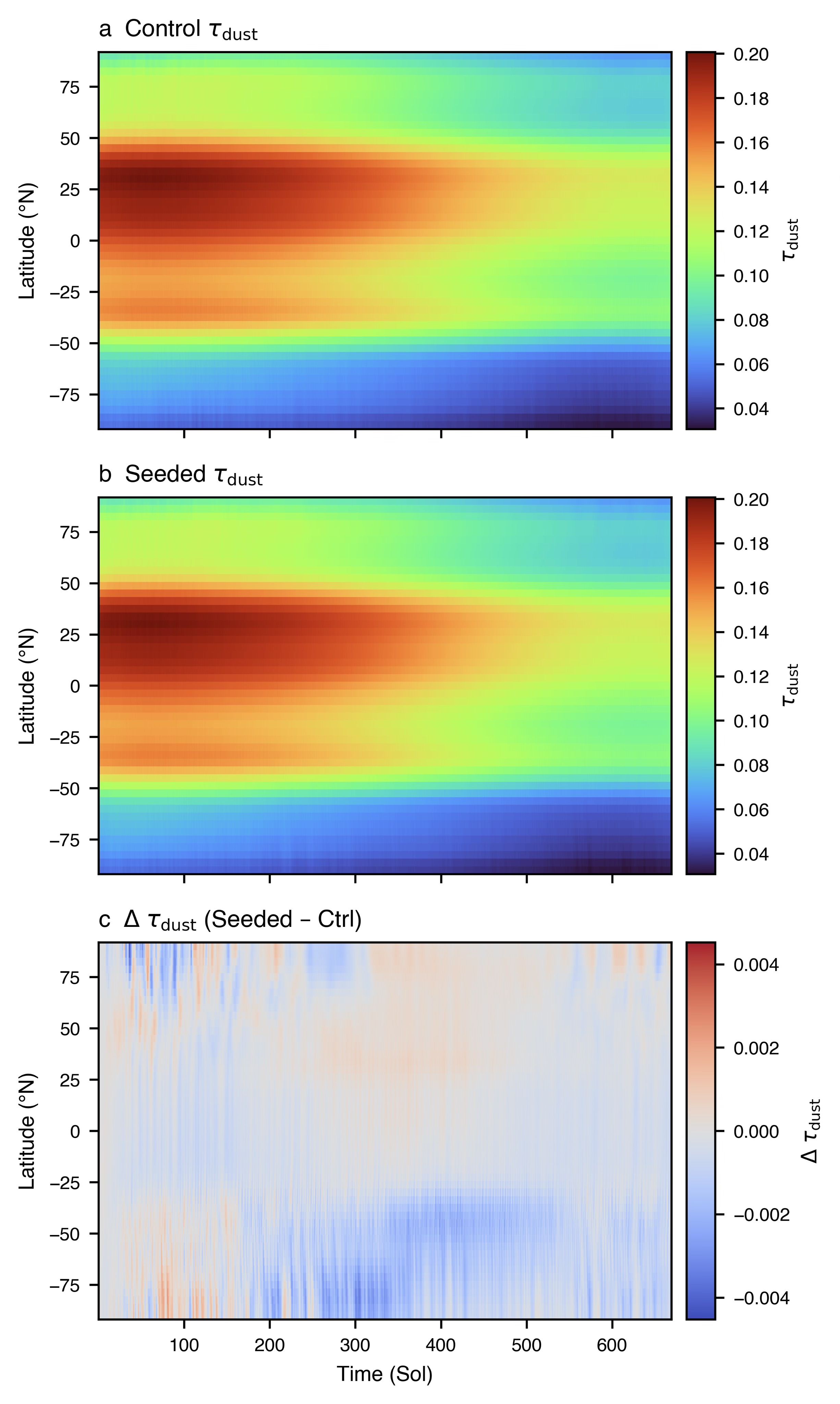}

\vspace{0.8em}
\textbf{Extended Data Figure 6.} Seasonal evolution of aerosol optical depth across 669 Sol

\textbf{(a)} Aerosol optical depth for the control simulation, showing the spatial and temporal distribution of the background aerosol \textbf{(b)} Aerosol optical depth under single-injection aluminium nanoparticle substitution \textbf{(c)} Aerosol optical-depth anomaly between the seeded and control simulations, showing the evolution and decay of the aerosol optical signal in the absence of continuous replenishment
\end{minipage}

\vfill
\begin{center}

\end{center}

\clearpage 

\begin{multicols}{2}

\noindent
\begin{minipage}{\linewidth}
\textbf{\large Extended Data Table 1}

\vspace{0.8em}

\resizebox{\linewidth}{!}{
\setlength{\tabcolsep}{3pt}
\begin{tabular}{lllll}
\toprule
Variable & Mean & Max & Min & Units \\
\midrule
tsurf & 191.801 & 304.0719 & 138.207 & K \\
temperature & 169.146 & 250.386 & 107.5675 & K \\
ALB & 0.1345 & 0.5500 & 0.0 &  \\
ps & 452.378 & 1030.4058 & 101.584 & Pa \\
h2o\_vap & 3.08E-07 & 1.300E-03 & -1.9884e-14 & kg/kg \\
co2\_ice\_surf & 223.301 & 2981.705 & -1.3877e-17 & kg m$^{-2}$ \\
evap\_surf\_flux & -3.080E-04 & 0.3743 & -8.277 & W m$^{-2}$ \\
h2o\_ice\_col & 6.151E-08 & 8.968E-06 & -1.7258e-19 & kg m$^{-2}$ \\
h2o\_ice\_surf & 42.700 & 3400.111 & -1.0842e-19 & kg m$^{-2}$ \\
tau\_col & 0.14135 & 2.9099 & 0.0333 &  \\
taudust & 0.11114 & 0.3217 & 0.0284 &  \\
\bottomrule
\end{tabular}
}

\vspace{0.3em}
{\footnotesize \textbf{Extended Data Table 1.} Mean, max and min values over 669 sols under single-injection nanoparticle infusion.}

\vspace{1.0em}

\noindent

\textbf{\large Extended Data Table 2}

\vspace{0.8em}

\resizebox{\linewidth}{!}{
\setlength{\tabcolsep}{3pt}
\begin{tabular}{lllll}
\toprule
Variable & Mean & Max & Min & Units \\
\midrule
tsurf & 194.893 & 313.223 & 138.313 & K \\
temperature & 176.157 & 273.294 & 110.193 & K \\
ALB & 0.1375 & 0.5500 & 0.0 &  \\
ps & 515.775 & 1119.494 & 103.543 & Pa \\
h2o\_vap & 8.8161e-05 & 0.2068 & -0.0382 & kg/kg \\
co2\_ice\_surf & 180.378 & 2991.9233 & -1.3877e-17 & kg m$^{-2}$ \\
evap\_surf\_flux & -0.0578 & 169.117 & -109.2219 & W m$^{-2}$ \\
h2o\_ice\_col & 1.8865e-06 & 0.0041 & -1.71997e-18 & kg m$^{-2}$ \\
h2o\_ice\_surf & 42.665 & 3400.325 & -2.7755e-17 & kg m$^{-2}$ \\
tau\_col & 0.1966 & 6.1867 & 0.0339 &  \\
taudust & 0.1643 & 0.7728 & 0.0296 &  \\
\bottomrule
\end{tabular}
}

\vspace{0.8em}
{\footnotesize \textbf{Extended Data Table 2.} Mean, max and min values over 669 sols for the control run.}
\end{minipage}

\vspace{2em}

\columnbreak

\noindent\textbf{\large Acknowledgments}

\vspace{0.5em}
\noindent This research was supported by the Excellence Cluster ORIGINS which is funded by the Deutsche Forschungsgemeinschaft (DFG, German Research Foundation) under Germany's Excellence Strategy - EXC-2094-390783311. The simulations have been carried out on the computing facilities of the Computational Center for Particle and Astrophysics (C2PAP) and of the Leibniz Supercomputer Center (LRZ). We like to thank Apl. Prof. Dr. Klaus Dolag for facilitating access to the computing resources at LRZ.

\vspace{0.8em}

\noindent\textbf{\large Declaration of competing interest}

\vspace{0.5em}
\noindent The authors declare that they have no known competing financial interests or personal relationships that could have appeared to influence the work reported in this paper.

\vspace{0.8em}

\noindent\textbf{\large References}

\vspace{0.8em}
\noindent Kerber, L., Forget, F. and Wordsworth, R. (2015) 'Sulfur in the early martian atmosphere revisited: Experiments with a 3-D Global Climate Model', \textit{Icarus}, 261, pp. 133--148.

\vspace{1em}
\noindent Wordsworth, R., Forget, F., Millour, E., Head, J.W., Madeleine, J.-B. and Charnay, B. (2013) 'Global modelling of the early Martian climate under a denser CO2 atmosphere: Water cycle and ice evolution', \textit{Icarus}, 222(1), pp. 1--19.

\vspace{1em}
\noindent Wordsworth, R.D., Kerber, L., Pierrehumbert, R.T., Forget, F. and Head, J.W. (2015) 'Comparison of ``warm and wet'' and ``cold and icy'' scenarios for early Mars in a 3-D climate model', \textit{Journal of Geophysical Research: Planets}, 120(6), pp. 1201--1219.

\vspace{1em}
\noindent Urata, R.A. and Toon, O.B. (2013) 'Simulations of the martian hydrologic cycle with a general circulation model: Implications for the ancient martian climate', \textit{Icarus}, 226(1), pp. 229--250.

\vfill
\end{multicols}

\newpage

\begin{multicols}{2}

\noindent Wordsworth, R., Kalugina, Y., Lokshtanov, S., Vigasin, A., Ehlmann, B., Head, J., Sanders, C. and Wang, H. (2017) 'Transient reducing greenhouse warming on early Mars', \textit{Geophysical Research Letters}, 44(2), pp. 665--671.

\vspace{1.2em}
\noindent Simonetti, P., Vladilo, G. et al. (2023) 'Seasonal thaws and liquid water reservoirs on Early Mars'.

\vspace{1.2em}
\noindent Ansari, S., Kite, E.S., Ramirez, R., Steele, L.J. and Mohseni, H. (2024) 'Feasibility of keeping Mars warm with nanoparticles', \textit{Science Advances}, 10(32), eadn4650.

\vspace{1.2em}
\noindent Bhatnagar, S., Codron, F., Millour, E., Bolmont, E., Brunetti, M., Kasparian, J., Turbet, M. and Chaverot, G. (2026) 'A Fast and Physically Grounded Ocean Model for GCMs: The Dynamical Slab Ocean Model of the Generic-PCM (rev. 3423)', \textit{Geoscientific Model Development}, pp. 1--41.

\vspace{1.2em}
\noindent Ojha, L., Karunatillake, S., Karimi, S. and Buffo, J. (2020) 'Groundwater production from geothermal heating on early Mars and implication for early martian habitability', \textit{Science Advances}, 6(48), eabb5360.

\vspace{1em}
\noindent Chaufray, J.-Y., Gonzalez-Galindo, F., Modolo, R., Leblanc, F., Forget, F., Liu, J., Millour, E., Lopez-Valverde, M., Steichen, V. and Chanteur, G. (2026) 'Helium on Mars', \textit{Icarus}, 453, 117058.

\vspace{1.2em}
\noindent Jakosky, B.M. and Edwards, C.S. (2018) 'Inventory of CO2 available for terraforming Mars', \textit{Nature Astronomy}, 2(8), pp. 634--639.

\vspace{1.2em}
\noindent Lange, L. and Forget, F. (2026) 'On the Possibility of Melting Water Ice During the Recent Past of Mars: Implications for the Recent Water Cycle', \textit{Journal of Geophysical Research: Planets}, 131, e2025JE009512.

\vspace{1.2em}
\noindent Verseux, C., Gargaud, M., Lehto, K. and Viso, M. (Eds.) (2024) \textit{Mars and the Earthlings: A Realistic View on Mars Exploration and Settlement}. Cham: Springer Nature Switzerland AG.

\vspace{1.2em}
\noindent Simonetti, P., Vladilo, G. et al. (2023) 'Seasonal thaws under mid-to-low pressure atmospheres on Early Mars'.

\vspace{1.2em}
\noindent Maurice, M., Forget, F., Lefèvre, F., Stolzenbach, A., Jaziri, A.Y., Braude, A.S., Turbet, M., Millour, E. and Luo, Y. (2026) 'Not enough H2O2 to warm early Mars', \textit{Icarus}, 448, 116914.

\vspace{1.2em}
\noindent Gauvain, A. (2025) 'A Global High-Resolution Hydrological Model to Simulate the Dynamics of Surface Liquid Reservoirs: Application on Mars', \textit{Earth Surface Dynamics / Zenodo}, doi:10.5281/zenodo.17208793.

\vspace{1.2em}
\noindent Haberle, R.M., Clancy, R.T., Forget, F., Smith, M.D. and Zurek, R.W. (Eds.) (2017) \textit{The Atmosphere and Climate of Mars}. Cambridge: Cambridge University Press.

\vspace{1.2em}
\noindent Morgan, G.A., Putzig, N.E., Baker, D.M.H. et al. (2025) 'Refined Mapping of Subsurface Water Ice on Mars to Support Future Missions', \textit{The Planetary Science Journal}, 6, 29.

\vfill
\end{multicols}

\newpage

\begin{multicols}{2}

\noindent Brines, A., López-Valverde, M.A., González-Galindo, F. and Funke, B. (2025) 'Water vapour vertical distribution on Mars after six years of NOMAD/TGO solar occultation observations', \textit{Journal of Geophysical Research: Planets}, 131, e2024JE008916.

\vspace{1.2em}
\noindent Wordsworth, R.D. (2016) 'The climate of early Mars', \textit{Annual Review of Earth and Planetary Sciences}, 44, pp. 381--408.

\vspace{1.2em}
\noindent Forget, F., Wordsworth, R., Millour, E., Madeleine, J.-B., Kerber, L., Leconte, J., Marcq, E. and Haberle, R.M. (2013) '3D modelling of the early martian climate under a denser CO2 atmosphere: Temperatures and CO2 ice clouds', \textit{Icarus}, 222(1), pp. 81--99.

\vspace{1.2em}
\noindent DeBenedictis, E.A. et al. (2025) 'Engineering planetary climates and habitability', \textit{Nature Astronomy}, 9, pp. 510--522.

\vspace{1.2em}
\noindent Turbet, M., Forget, F., Leconte, J., Charnay, B. and Tobie, G. (2019) 'CO2 condensation is a serious limit to the deglaciation of Earth-like planets', \textit{Scientific Reports}, 9, 42030.

\vfill
\columnbreak
\null
\vfill
\end{multicols}

\end{document}